\documentstyle[13lomcon,cite,graphicx]{article}

\begin{document}

\title{STRUCTURE FUNCTION MOMENTS OF PROTON AND NEUTRON}

\author{ M.Osipenko \footnote{e-mail: osipenko@ge.infn.it}}

\address{INFN, Sezione di Genova, 16146 Genova, Italy, \\
Moscow State University, Skobeltsyn Institute of Nuclear Physics, 119992 Moscow, Russia}


\maketitle\abstracts{ QCD-inspired phenomenological analysis of experimental moments
of proton and deuteron structure functions $F_2$ have been presented. The obtained
results on the d/u ratio at large-$x$, isospin dependence of higher twists and
comparison with Lattice QCD calculations were discussed. We remind shortly these results:
the obtained ratio is consistent with the asymptotic limit $d/u\to 0$ at $x\to 1$,
the total contribution of higher twists is found to be isospin independent and
the non-singlet moments are in excellent agreement with the Lattice data.
We present here some details of the analysis triggered by the public discussion.}

Measurements of the nucleon structure function $F_2$ provide the information about
the longitudinal momentum distribution of partons. These distributions
being governed by soft strong interactions cannot be described by perturbative
QCD methods. Only Lattice QCD simulations allow to evaluate these quantities.
Recent measurements of proton and deuteron structure function moments
over wide $Q^2$-interval~\cite{f2p_paper,f2d_paper}
and the evaluation of neutron moments~\cite{f2n_paper}
allowed to improve the knowledge of these non-perturbative distributions.
Detailed descriptions of these analyses are given in papers mentioned
above, whereas in the present proceeding we develop further two arguments
selected by the public discussion.

\begin{figure}[!h]
\begin{center}
\includegraphics[bb=2cm 6cm 20cm 22cm, scale=0.35]{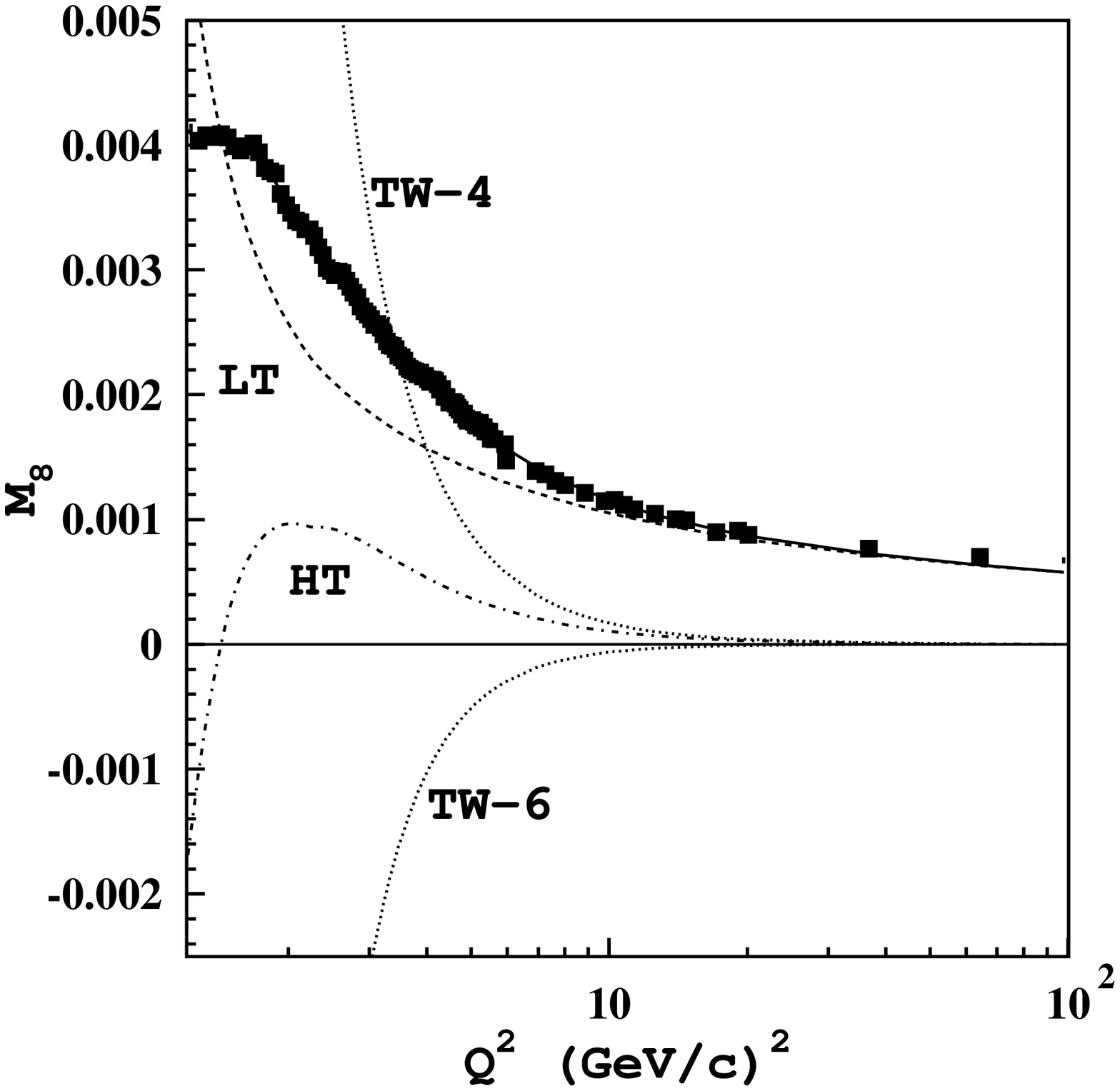}%~~~
\includegraphics[bb=2cm 6cm 20cm 22cm, scale=0.35]{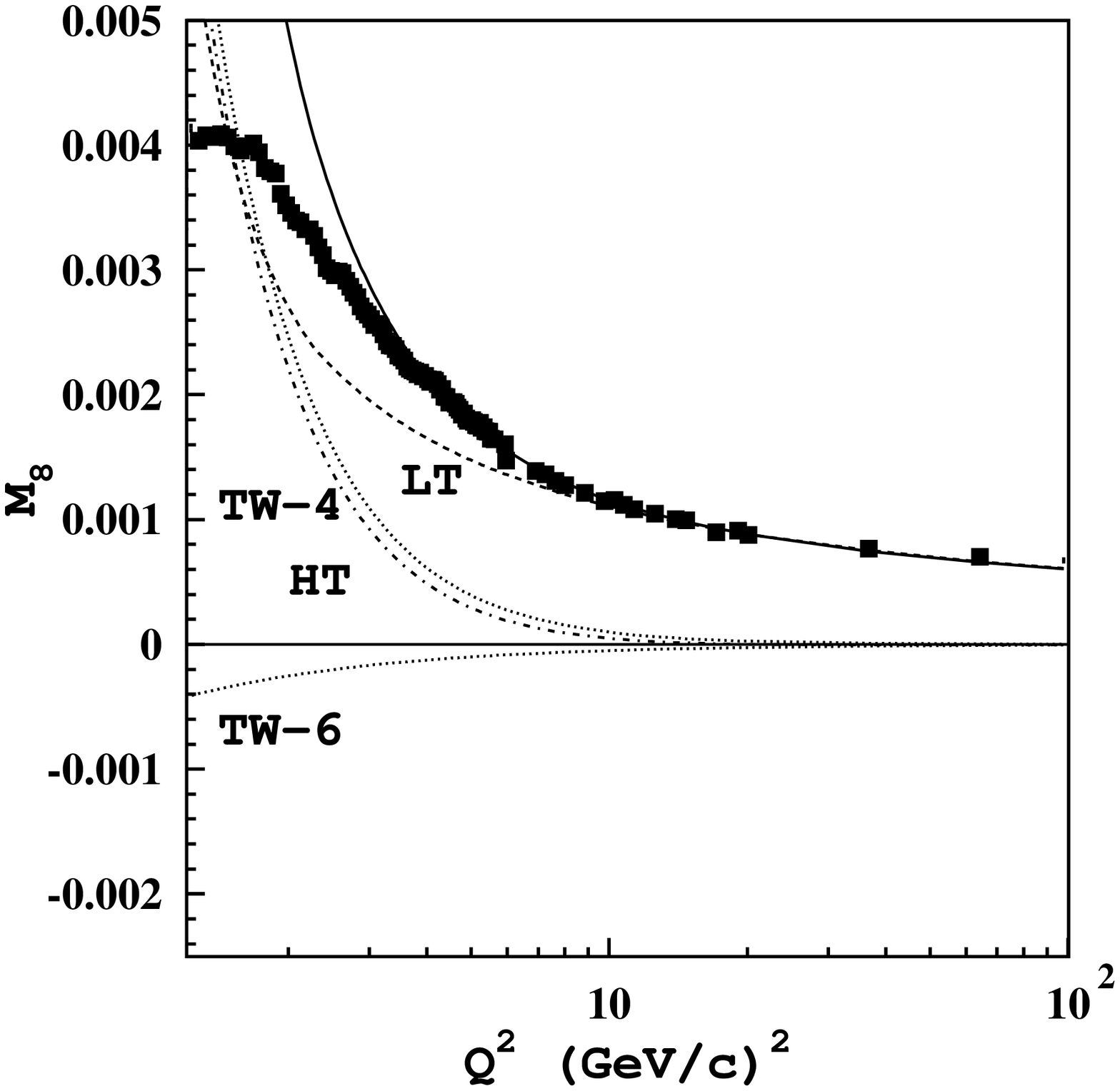}
\caption{\label{fig:anomal_dim} Fit of the structure function moment $M_8$
with Eq.~\ref{eq:ope} using higher twist anomalous dimensions $\gamma_n^\tau$
as free parameters (left) and assuming $\gamma_n^\tau=0$ (right):
dashed line - the leading twist contribution,
dotted lines - twist-4 and twist-6 contributions,
dot-dashed line - the total higher twist contribution,
solid line - the total fit.}
\end{center}
\end{figure}

%%%%%%%%%%pQCD part%%%%%%%%%%%%%%%%%%%%%%
Experimentally extracted moments of the proton and deuteron structure
functions $F_2$ were analyzed to separate leading twist (LT) and higher twist (HT) terms.
This was performed by fitting the data $Q^2$-dependence with the following expression:
\begin{equation}\label{eq:ope}
M_n (Q^2)= \int_0^1 dx x^{n-2} F_2(x,Q^2) = LT_n(\alpha_S)
+ \sum_{\tau=4}^k a_n^\tau
\Biggl(\frac{\alpha_S(Q^2)}{\alpha_S(\mu^2)}\Biggr)^{\frac{\gamma_n^\tau}{2\beta_0}}
\Biggl(\frac{\mu^2}{Q^2}\Biggr)^{\frac{\tau-2}{2}} ~,
\end{equation}
\noindent where $LT_n$ is the LT part of the $n$-th moment evaluated at NLL accuracy,
$\alpha_S$ is the running coupling constant,
$\mu^2$ is an arbitrary scale (taken to be 10 (GeV/c)$^2$),
$a_n^\tau$ is the matrix element of corresponding QCD operator,
$\gamma_n^\tau$ is its anomalous dimension,
$\beta_0=11-\frac{2}{3}N_F$ with $N_F$ being number of active flavors,
$\tau$ is the order of the twist and $k$ is the maximum HT order considered.
The number of HT terms ($k$) in the expansion~\ref{eq:ope} is, of course,
arbitrary because we don't know at which $1/Q^2$ power
the series converges.
Moreover, anomalous dimensions of perturbative coefficients
in front of HT terms are known in a very few cases~\cite{Shuryak,Kawamura}.
Most of $x$-space analyses neglect this dependence
assuming $\gamma_n^\tau=0$ for $\tau>2$. In the presented analyses the anomalous dimensions
were varied as free parameters and extracted from the best fit to the data. The results show
a very strong sensitivity of the fit to the values of HT anomalous dimensions at low-$Q^2$.
Indeed, it can be seen in the comparison of two twist expansions shown in Fig.~\ref{fig:anomal_dim}:
one using HT anomalous dimensions as free parameters and another one assuming them to be zero.
The lower limit of the fitted $Q^2$-interval was taken
to be 1.2 (GeV/c)$^2$ for the full fit. In the fit with fixed anomalous dimensions
it was increased to 3.6 (GeV/c)$^2$ by the condition
of having the same $\chi^2$ per number of degrees of freedom.
It is evident that only the variation of anomalous dimensions
permits to describe the data until $Q^2=1.2$ (GeV/c)$^2$ by two HT terms.
This observation emphasizes that the knowledge of perturbative anomalous
dimensions of HT terms is crucial to single out individual HT operator matrix elements.

\begin{figure}[!h]
\begin{center}
\includegraphics[bb=5cm 8cm 15cm 20cm, scale=0.4,angle=-90]{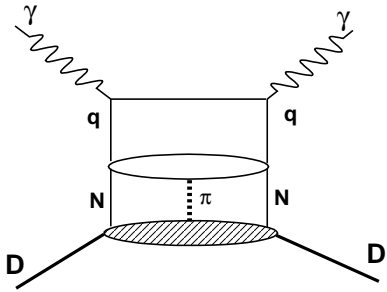}
\caption{\label{fig:fsi_diag} Example of FSI mechanism in the inclusive
electron-deuteron scattering. The nuclear interaction between the nucleon
spectator and nuclear medium is likely carried by a colorless pion.}
\end{center}
\end{figure}

%%%%%%%%%%%Nuclear part%%%%%%%%%%%%%%%
In the extraction of neutron moments we assumed the dominance of the Impulse Approximation (IA)
in the LT part of deuteron moments and treated other nuclear effects as, model dependent,
corrections to this approximation. This allowed for a simple
extraction of LT neutron moments from the following algebraic relation:
\begin{equation}\label{eq:nucl_cor}
M_n^n(Q^2)=\frac{2M_n^D(Q^2)}{N_n^D}-M_n^p(Q^2) ~,
\end{equation}
\noindent where $M_n^p$, $M_n^n$ and $M_n^D$ are LT moments of the proton, neutron and deuteron,
respectively. $N_n^D$ is the moment of the nuclear momentum distribution $f^D$
i.e. the structure function of the deuteron composed of point-like nucleons
(see Ref.~\cite{f2n_paper} for details).
The dominance of IA in LT moments, however, implies that processes beyond IA contribute mainly to HT terms.
These processes are the scattering off correlated nucleons (Final State Interaction (FSI)) and
the scattering off a nuclear constituent different than the nucleon (Meson Exchange Current (MEC)).
Only the lowest $n=2$ LT moment, sensitive to the low-$x$ dynamics,
carries a small contribution from FSI and MECs estimated to be $<0.5$ \%,
whereas it is found to be negligible for higher ($n>2$) LT moments.
In fact, the LT part of FSI shown in Fig.\ref{fig:fsi_diag} contaminates structure functions
at very low $x$ ($x<0.1$) values because the nucleon spectator has
a long time $\xi^0 \leq 1/Mx$ (here $M$ is the nucleon mass) to interact with nuclear environment
while awaiting return of the active quark\cite{JAFFE}. However, for higher moments ($n>2$)
the mean $x$ value is close to unity and therefore the time left for the interaction
$\xi^0 \leq 1/M << 1/m_{\pi}$. Here we assume that nuclear interactions are carried mostly by pions.

\begin{figure}[!h]
\begin{center}
\includegraphics[bb=2cm 6cm 22cm 23cm, scale=0.4]{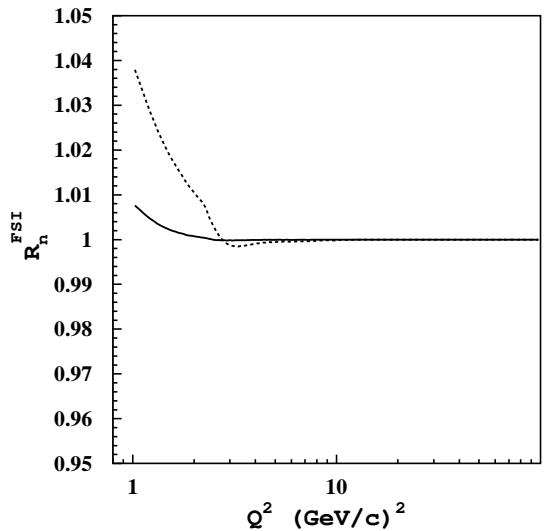}
\caption{\label{fig:fsi_mom} Ratio of the deuteron structure function
moments calculated using the parameterization of $F_2^D$ from Ref.~\cite{Simula_qe}
including and excluding FSI in the quasi-elastic channel:
the solid line - $n=2$,
the dashed line - $n=8$.}
\end{center}
\end{figure}

Hence, the bulk of FSI was expected to give a contribution to the HT term of the moment
expansion, where the current quark rescatters from the nuclear spectator.
In order to test this assumption phenomenologically we used calculations of FSI
in the quasi-elastic peak region based on the model from Ref.~\cite{Simula_qe}.
To this end, we computed moments of the modeled deuteron structure function $F_2$
including and excluding FSI. The ratios between these two calculations for $n=2$ and $n=8$
are shown in Fig.~\ref{fig:fsi_mom}. From the figure it follows that, indeed,
FSI contribution disappears at large $Q^2$ generating an additional HT term,
while LT part is unaffected by this contribution.
Moreover, the size of FSI contribution in moments does not exceed few \%
at lowest analyzed $Q^2$, and it is much smaller than the total nucleon HT contribution
estimated to be about $\approx 25$\% at the same $Q^2$. We speculate, therefore,
that also the HT term has not more than 20\% contamination from FSI,
and Eq.~\ref{eq:nucl_cor} is also applied to this term within $\approx 20$\% accuracy.
This accuracy is comparable to the precision of the extracted total HT term~\cite{f2p_paper,f2d_paper}.

%%%%%%%%%%Summary%%%%%%%%%%%%%%%%%%%%%%
Summarizing, the presented analysis of the experimental moments
of proton and deuteron structure functions $F_2$ showed that:
\begin{itemize}
\item
knowledge of perturbative anomalous dimensions of higher twist terms
is crucial to single out individual higher twist operator matrix elements;
\item
for $n>2$ FSI mechanism appears in the nuclear structure function moments
as an additional higher twist term. Its partial estimates indicate that
the relative contribution of FSI to the total higher twist term does not exceed 20\%,
comparable to the precision of the total higher twist extraction.
\end{itemize}

\end{document}